# The Density of Iterated Crossing Points and a Gap Result for Triangulations of Finite Point Sets

Rolf Klein *
Institute of Computer Science I
D-53117 Bonn, Germany
rolf.klein@uni-bonn.de

Martin Kutz
Max-Planck-Institut für Informatik
66123 Saarbrücken, Germany
mkutz@mpi-inf.mpg.de

**Abstract**

Consider a plane graph $G$, drawn with straight lines. For every pair $a, b$ of vertices of $G$, we compare the shortest-path distance between $a$ and $b$ in $G$ (with Euclidean edge lengths) to their actual distance in the plane. The worst-case ratio of these two values, for all pairs of points, is called the *dilation* of $G$. All finite plane graphs of dilation 1 have been classified. They are closely related to the following iterative procedure. For a given point set $P \subseteq \mathbb{R}^2$, we connect every pair of points in $P$ by a line segment and then add to $P$ all those points where two such line segments cross. Repeating this process infinitely often, yields a limit point set $P^\infty \supseteq P$. This limit set $P^\infty$ is finite if and only if $P$ is contained in the vertex set of a triangulation of dilation 1.

The main result of this paper is the following gap theorem: For any finite point set $P$ in the plane for which $P^\infty$ is infinite, there exists a threshold $\lambda > 1$ such that $P$ is not contained in the vertex set of any finite plane graph of dilation at most $\lambda$. As a first ingredient to our proof, we show that such an infinite $P^\infty$ must lie dense in a certain region of the plane. In the second, more difficult part, we then construct a concrete point set $P_0$ such that any planar graph that contains this set amongst its vertices must have a dilation larger than 1.0000047.

## 1 Introduction

Let us consider the following scenario described by D. Eppstein in [10]. A typical university campus contains facilities like lecture halls, dorms, library, mensa, and supermarkets, which are connected by some path system. Students in a hurry are tempted to walk straight across the lawn, if the shortcut seems worth it. After a while, this causes new paths to appear. Since their intersections are frequented by many people, they attract coffee shops or other new facilities. Now, people will walk across the lawn to get quickly to a coffee shop, and so on. A natural question discussed in [10] is

**(A)** What happens if this process of forming intersections and treading new paths continues?

Eppstein gives a qualitative answer by completely classifying all those finite point sets in the plane for which this process comes to a halt. In this paper we strengthen that result by showing that, otherwise, not only will infinitely many new intersections be created but these intersections will also lie dense in a region of positive measure. So actually, part of the lawn will completely be gone.

*This work was partially supported by the German Research Foundation DFG, grant no. Kl 655/ 14-1, 14-2.

In the unfortunate case that the process would not come to a halt for an initial set of sites, we now assume that students could be convinced to stay on the official paths if the resulting detours, as compared to straight-line walks, were sufficiently small. In this paper we address this relaxation of the exact problem above, which has been brought up recently by Ebbers-Baumann et al. [8].

**(B)** Could the original campus map be designed in such a way that the process never starts because the advantage of walking straight is always negligible?

The main result of this paper is a negative answer to Question B. For a formal statement, a few precise definitions are indispensable.

**Definition 1.** For a point set $P \subseteq \mathbb{R}^2$ let $\mathcal{L}(P)$ denote the set of all line segments between pairs of points in $P$. We define the *complete intersection set* $\dot{P}$ of $P$ to be the union of $P$ with the set of crossings points of lines in $\mathcal{L}(P)$; formally, $\dot{P} = P \cup \bigcup\{\ell_1 \cap \ell_2 \mid \ell_1, \ell_2 \in \mathcal{L}(P) \text{ and } \ell_1, \ell_2 \text{ not parallel}\}$. Then define $P^k$ for all $k \geq 0$ by $P^0 = P$ and $P^{k+1} = (P^k)^{\dot{}}$, and finally let $P^\infty := \bigcup_{k\geq 0} P^k$. We call a finite point set $P$ *stable* if $\dot{P} = P$ holds, and *stabilizing* if $P^\infty$ is finite.

We have to exclude intersections of parallel segments in the above definition for the simple reason that such overlaps would give one-dimensional intersections and not discrete points. This choice corresponds well with our campus scenario: four locations on a straight line should not create any new "crossings" between them.

We can now make our strengthened answer to Question A precise.

**Theorem 1.** *For every non-stabilizing finite point set $P \subset \mathbb{R}^2$, the limit set $P^\infty$ lies dense in some region of positive measure.*

In order to capture the relaxed version in Question B, we need the concept of the *dilation* of a plane graph.

**Definition 2.** Let $G = (V, E)$ be a plane graph, i.e., its vertices are mapped to points in the plane and the edges are mapped to connecting curves that pairwise do not cross. The *dilation* of $G$ is defined as

$$\delta(G) := \max_{p \neq q \in V} \frac{d_G(p, q)}{|pq|}, \tag{1}$$

where $d_G(p, q)$ denotes the length of a shortest path in $G$ connecting $p$ and $q$, and $|pq|$ is the Euclidean distance.

There are obvious connections between stable and stabilizing point sets and the dilation of plane graphs.

**Lemma 1.** *If a finite point set $V$ is stable then there exists a unique maximal triangulation $T$ of $V$ and this $T$ is a graph of dilation 1. Converseley, every dilation-1 graph is a triangulation and its vertex set is stable. Exactly the subsets of finite stable sets are stabilizing.*

The following definition provides the formal notion behind Question B.

**Definition 3.** Let $P$ be a finite set of points in the plane. The *dilation*, $\Delta(P)$, of $P$ is defined as

$$\Delta(P) := \inf\{\delta(T) \mid T = (V, E) \text{ finite triangulation with } P \subseteq V\}.$$

We should remark that the dilation of a point set is invariant under scaling because the quotient in (1) is. Our main result now reads:

**Theorem 2.** *Every non-stabilizing finite point set in the plane has dilation strictly greater than 1.*

In other words, this gap theorem states that if a point set $P$ is not one of the few nicely behaved ones then there exists a threshold $\lambda > 1$ such that $P$ cannot be contained in any finite plane graph of dilation smaller than $\lambda$.

## 1.1 Related work

The problem of how to connect a given point set by a network of small dilation has been extensively studied in the context of spanners, where the dilation of a graph is also called its stretch factor or spanning ratio. There exists a wealth of results on constructing spanners with a dilation arbitrarily close to 1, that have other nice properties like, e. g., linear size, a weight not much exceeding that of the minimum spanning tree, or bounded degree; see the monographs [9] and [18]. But these spanners can contain many edge crossings.

If crossings are not allowed (since bridges are too expensive) one could ask for the plane graph of lowest possible dilation over the given point set, $P$. Such a graph can be assumed to be a triangulation of $P$, because curved edges could be pulled taut and adding non-crossing edges cannot increase the dilation. In his handbook chapter [9], Eppstein has posed the following open problems.

- (Problem 8) Can the minimum dilation triangulation of a given point set be computed in polynomial time?
- (Problem 9) How large a dilation can the minimum dilation triangulation have, in the worst case?

In Problem 9, an upper bound is the dilation of the Delaunay triangulation, which has been shown to be less than 2.42 by Keil and Gutwin [14], while a tight bound of $\pi/2$ is conjectured. A lower bound is $\sqrt{2}$, attained by the vertices of a square. A partial answer to Problem 8 was recently given by Eppstein and Wortman [11] who provided a randomized $O(n \log n)$ algorithm for computing an optimal star center for $n$ points in $d$-space.

The problem setting we consider in this paper can be seen as the "Steiner point" variant of Problem 9 above: Instead of looking for an optimal triangulation on just the given point set, we allow ourselves the introduction of a finite number of auxiliary "Steiner points" to help reduce the overall dilation. The crucial point here is, of course, that dilation amongst the new points, which we introduce to reduce detours for the given vertices, also needs to be controlled. Theorem 2 shows that essentially every point configuration has an intrinsic irregularity that cannot be smoothed out. Ebbers-Baumann et al. [8] have studied Problem 9 in this model. They provided an upper bound of 1.1247 but left open the question of a non-trivial lower bound for finite point sets. The present paper fills this gap.

We have already emphasized that a point set is stabilizing if and only if it is contained in the vertex set of a triangulation of dilation 1. Figure 1 shows the complete classification of such triangulations from [10]. There are two infinite families and one exceptional case.

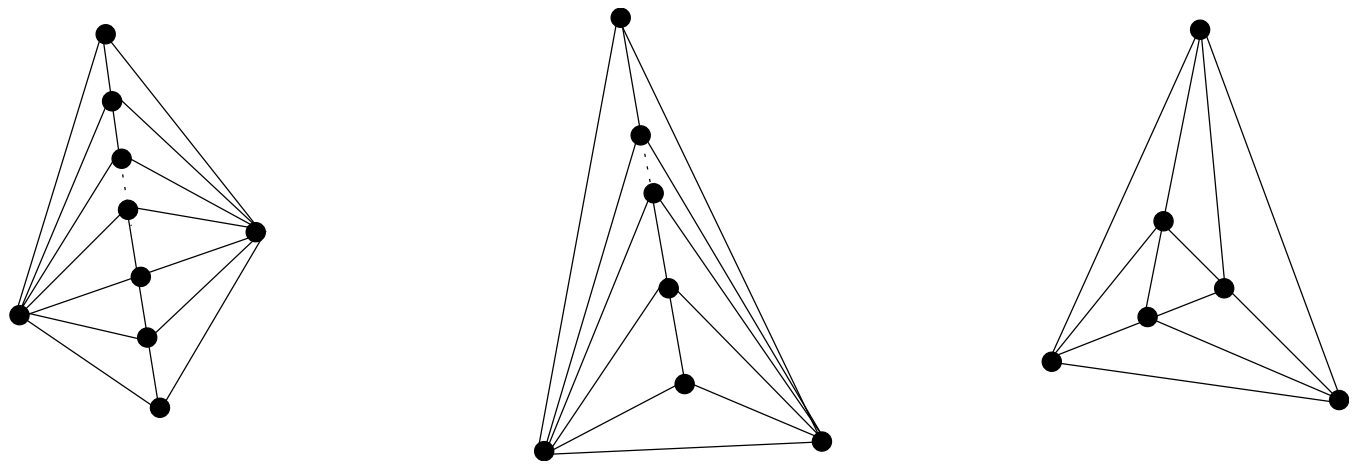

Figure 1: The triangulations of dilation 1.

Quite recently, a related measure called *geometric dilation* has been introduced, see [1, 3, 4, 5, 6, 7], where *all* points of the graph, vertices and interior edge points alike, are considered in computing the dilation. The small difference in definition leads to rather different results. For example, plane graphs of minimum geometric dilation tend to have curved edges. Yet, it is interesting to observe that in this model non-trivial upper and lower bounds could be established. Each finite point set can be embedded into a plane graph of geometric dilation $\leq 1.678$, and there are point sets for which each plane graph containing them has a geometric dilation $\geq (1 + 10^{-11})\pi/2$; see [3, 7]. These results are based on methods from convex geometry that do not apply here.

For a point set $P$ in general position we can expect a biquadratic number of crossings in the complete graph $C(P)$, due to the Crossing-Number Theorem; see, e. g., Matoušek [16], but this result does not imply where the crossing points are located. Bezdek and Pach [2] have studied iterated intersections of unit circles. Ismailescu and Radoičić [12] have recently shown that the iterated construction of the intersections of all *lines* through the point pairs of a given set $P$ is dense in the whole plane, except for two special cases. Since their methods heavily use intersection points that lie outside the connecting line segments, we cannot adapt them to our problem. In fact, we shall see that our Theorem 1 is actually a stronger statement and implies the result of [12].

### 1.2 Overview

The rest of this paper is organized as follows. In Section 2 we prove that for each set $P$ of five points in convex position, their limit set $P^\infty$ lies dense in the convex hull of $P^1$; see Figure 2. This result has independently been shown by Kamali [13] by different methods. As a consequence, we obtain the density result stated in Theorem 1. This also yields, as a byproduct, a new simple proof for the density result on iterated line intersections from [12].

In Section 3 we consider triangulations whose dilation is almost equal to 1, and show that they contain good approximations of the complete intersection sets $P^k$, for any finite subset $P$ of vertices, and for each fixed order $k$. Those results form the bridge between Theorem 1 about dense iterated intersections and Theorem 2 about point sets with dilation greater than 1. The concept of "approximations" of point sets will be required frequently throughout this paper, so let us introduce a formal notion.

**Definition 4.** We say that a set $Q \subseteq \mathbb{R}^2$ is an *$\epsilon$-cover* of a set $P \subseteq \mathbb{R}^2$ if the (closed) $\epsilon$-neighborhood of every point of $P$ contains a point of $Q$.

In other words, $Q$ is an $\epsilon$-cover for $P$ if the (directed) Hausdorff distance from $P$ to $Q$ is at most $\epsilon$.

In Section 4 we provide a lower bound to the Steiner-point version of Problem 9 above by constructing a point set of dilation larger than 1.0000047. Precisely, the set of 200 points evenly placed on the boundary $B$ of the unit square turns out to have this property (Theorem 5 and Corollary 1).

Our construction uses the following novel technique. We show that if some set $V$ is an $\epsilon$-cover for $B$ (for some sufficiently small $\epsilon$) then the point set $V^2$ approximates a scaled copy $B'$ of $B$ with a tolerance smaller than the scaling factor times $\epsilon$. If, in addition, $V$ is the vertex set of a triangulation of sufficiently small dilation, then, by the result obtained in Section 3, the set $V^2$, and therefore $B'$, are closely approximated by vertices in $V$. By infinite descent, this argument shows that $V$ cannot be finite.

Finally, we combine the above results to prove that for each point set $P$ that is not contained in the vertex set of one of the triangulations in Figure 1, there exists a bound $\lambda > 1$ such that each finite plane graph containing $P$ in its vertex set has dilation at least $\lambda$. We conclude with some open problems and suggestions for further research.

## 2 The Density of the Set of Iterated Crossing Points

We begin with the density results for dilation 1. Any set of four points in the plane is easily seen to be stabilizing. So let us consider five points in convex position and show that their limit is dense. The generalization to arbitrary non-stabilizing point sets will be easy.

**Theorem 3.** *For any set $P$ of five points in convex position, the limit $P^\infty$ lies dense in the region $A = \mathrm{conv}(P^1 \setminus P)$. (See Figure 2.) Precisely, for every $\epsilon > 0$, there exists an index $k$ such that the $\epsilon$-neighborhood of every point in $A$ contains a point from $P^k$, i.e., $P^k$ is an $\epsilon$-cover for $A$.*

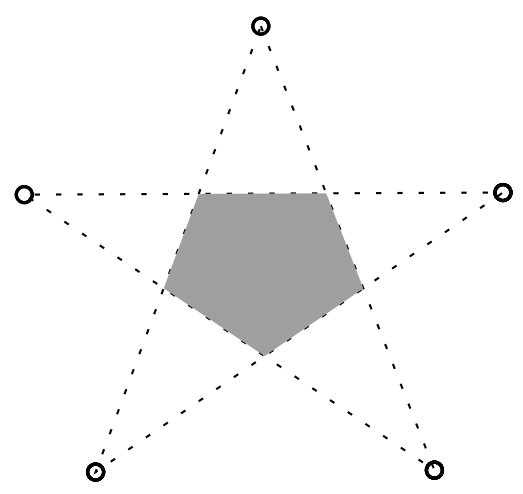

Figure 2: Five points in convex position with their dense limit shaded.

*Proof.* We show that some $P^k$ is an $\epsilon$-cover for the boundary of the region $A$. Then the claim for the interior of $A$ immediately follows.

Consider the configuration in the left of Figure 3. The two points $x$ and $y$ there are assumed to lie in some $P^j$ and so $\bar{x}$ and $\bar{y}$, their projections onto the boundary line $ab$, lie in $P^{j+1}$. We create a sequence of intersection points $x_i, y_i$ on the lines $x\bar{x}$ and $y\bar{y}$ by zig-zagging between $v$ and $w$ as shown in the picture. This way we obtain a pair $x_t, y_t$ arbitrary close to the line $ab$ so that the ratio $|x_t\bar{x}|/|y_t\bar{y}|$ gets arbitrarily close to $|\bar{x}w|/|\bar{y}w|$.

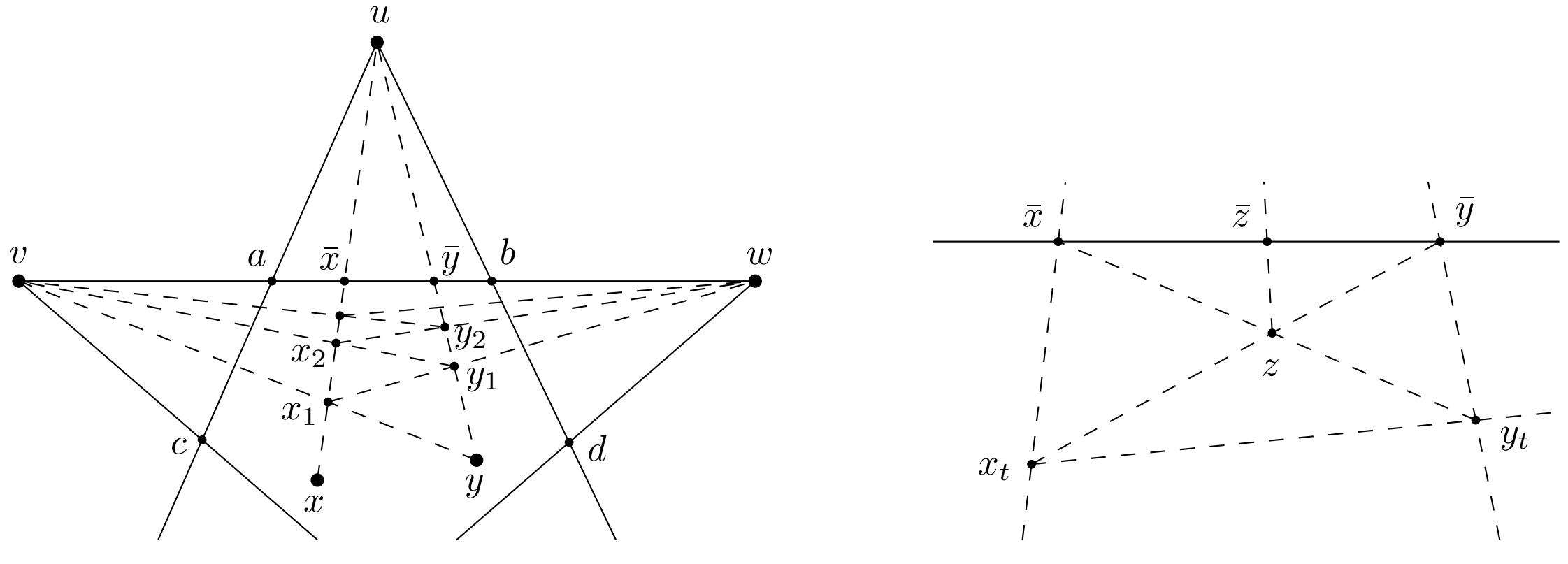

Figure 3: Covering the boundary of $A$.

Now we form the intersection $z$ of the diagonals $x_t\bar{y}$ and $y_t\bar{x}$ and the projection $\bar{z}$ onto the line $ab$, as shown in the right of Figure 3. Again up to arbitrarily small errors, we get

$$\frac{|\bar{z}\bar{x}|}{|\bar{z}\bar{y}|} \approx \frac{|x_t\bar{x}|}{|y_t\bar{y}|} \approx \frac{|\bar{x}w|}{|\bar{y}w|} < \frac{|aw|}{|bw|}.$$

So the ratio by which the new point $\bar{z}$ splits the segment $\bar{x}\bar{y}$ in two is bounded by a constant.

Note that the above construction needs the inner points $x$ and $y$ in order to create the subdivision $\bar{z}$ of the segment $\bar{x}\bar{y}$. But since the point $\bar{z}$ is created together with its interior partner $z$, we can repeat the construction on the segments $\bar{x}\bar{z}$ and $\bar{z}\bar{y}$. After recursing on the whole boundary of $A$ to a sufficiently fine resolution, we easily cover the interior of $A$ by intersections of connections between boundary points. □

Now the general density result of Theorem 1 follows from the easily verified fact that for any non-stabilizing point set $P$, the set $P^4$ must contain five points in convex position.

### 2.1 A related concept: exterior intersections

If the definition of the intersection set $\dot{P}$ was modified to include all intersections of infinite *lines* through pairs of points (instead of connecting segments, only), we would be able to create new points outside the convex hull of the initial set $P$. This setting was investigated by Ismailescu and Radoičić in [12]. They proved that except for two exceptional cases, all point sets grow densely into the whole plane.

**Theorem 4 (Ismailescu and Radoičić, 2004).** *The limit set of line intersections of a point set that contains four points, one of them in the interior of the convex hull of the others, is everywhere dense in the whole plane.*

We remark that our Theorem 3 directly implies this result, simply because the dense region $A$ in the pentagon can be used to shoot at any point in the plane with arbitrary precision. (The initial five-point configuration required for our result is easily obtained from the four points of Theorem 4 in a few steps.) This gives a very simple new proof for Theorem 4, which in [12] requires almost ten pages of geometric construction and algebraic argument.

## 3 Approximating Exact Intersections

The above density results rely crucially on knowing the exact intersection points as they are guaranteed by dilation 1. Next we show that if we allow for only slightly larger dilation, we can still get arbitrarily good approximations of the exact results.

This fact is based on the following observation. Let $T = (V, E)$ denote a triangulation of dilation $\delta$. By definition, each pair of vertices $p, q \in V$ is connected by a path $\pi(p, q)$ in $T$ whose length does not exceed $\delta \cdot |pq|$. In particular, for each vertex $t$ on $\pi$, we know that

$$\frac{|pt| + |tq|}{|pq|} \le \delta$$

holds. Thus, path $\pi$ is contained in the ellipse of diameter $\delta|pq|$, whose foci are $p$ and $q$; see Figure 4.[1]

The width of this ellipse is equal to $w = |pq|\ \sqrt{\delta^2 - 1}$; it tends to zero as the dilation tends to 1. This has a useful consequence. Suppose that vertices $p, w, q, v$ of $T$ are in convex position, and that dilation $\delta$ is sufficiently small. Then two shortest paths $\pi(p, q)$ and $\pi(v, w)$ in $T$ must cross at some vertex $t$ that is contained in the intersection of the ellipses. Clearly, the smaller dilation $\delta$ is, the closer must vertex $t$ be to the crossing point, $z$, of the segments $pq$ and $vw$.

**Lemma 2.** *For every finite point set $P \subseteq \mathbb{R}^2$, every natural number $k$, and any $\epsilon > 0$, there exists a bound $\delta > 1$ such that the following holds. If $T = (V, E)$ is triangulation of dilation $\delta(T) \le \delta$ such that $V$ contains $P$, then $V$ is an $\epsilon$-cover for $P^k$.*

[1] The *diameter* of an ellipse with focal points $p, q$ denotes the constant sum $|pb| + |bq|$, where $b$ is any point on the boundary.

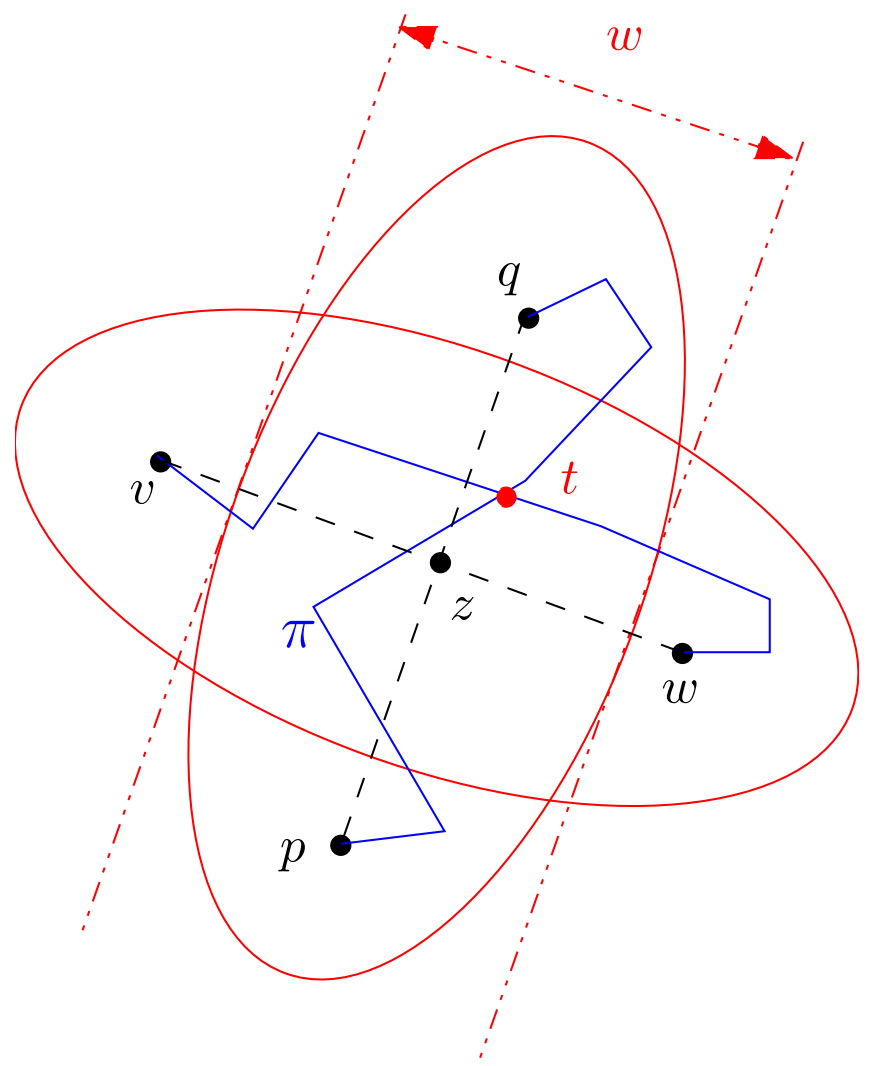

Figure 4: In a triangulation of dilation $\delta$, the shortest path between vertices $p$ and $q$ is contained in an ellipse of width $|pq|\sqrt{\delta^2-1}$.

*Proof.* Otherwise, there would be a sequence of triangulations $T_n = (V_n, E_n)$ of dilation $\delta(T_n) \leq 1 + \frac{1}{n}$ such that each vertex set $V_n$ contains $P$ but none of them is an $\epsilon$-cover for $P^k$. For each crossing point $a_i, 1 \leq i \leq t$, of $P^1$ we can find a sequence $(a_{i,n})_n$, where $a_{i,n} \in V_n$, such that $\lim_{n\to\infty} a_{i,n} = a_i$, by the ellipse argument from above. Now, let $b_j$ be a crossing point in $P^2$; we can find, in each complete intersection set of the set $\{a_{i,n}; 1 \leq i \leq t\} \subseteq V_n$, an element $b'_{j,n}$ such that $\lim_{n\to\infty} b'_{j,n} = b_j$. By the same argument as before, each set $V_n$ contains some vertex $b_{j,n}$ near $b'_{j,n}$ such that $\lim_{n\to\infty} b_{j,n} = b_j$ holds, too. Since the total number of points in $\bigcup_{i=1}^{k} P^i = P^k$ is finite, this argument, repeated $k$ times, shows that some set $V_n$, of large enough index $n$, will contain a subset that is an $\epsilon$-cover for $P^k$—a contradiction. □

# 4 Constructing Point Sets of Dilation Larger than 1

Let us quickly recall the central definitions from the preceding sections. The *dilation,* $\Delta(P)$, of a finite point set $P$ is the infimum of the dilations of all finite triangulations containing $P$ amongst their vertices. We say that $Q$ is an *$\epsilon$-cover* for some point set $P$ if each point of $P$ has a point of $Q$ in its $\epsilon$-neighborhood. Finally, $P^k$ denotes the set of all crossing points that result from iterating the construction of the complete set of intersections $k$ times, starting with point set $P$.

The goal of this section is in providing concrete point sets $Q$ whose dilation $\Delta(Q)$ can be bounded away from 1. The following main lemma provides the central tool.

**Lemma 3.** *There are values $0 < a < A$, $0 < \epsilon < a/2$, and $1 < \delta$, such that the following holds. Let $V$ be a point set such that*

1. *$V$ is an $\epsilon$-cover for the boundary $B$ of a square of side length $A + a$ and*

2. *$V$ is vertex set of some triangulation $T$ of dilation $\leq \delta$.*

*Then $V$ is an $\frac{A-a}{A+a} \cdot \epsilon$-cover for the boundary $B'$ of a square of side length $A - a$, centered in $B$.*

*Proof (sketch).* Here we present the ideas behind the proof; the technical details are given in the Appendix.

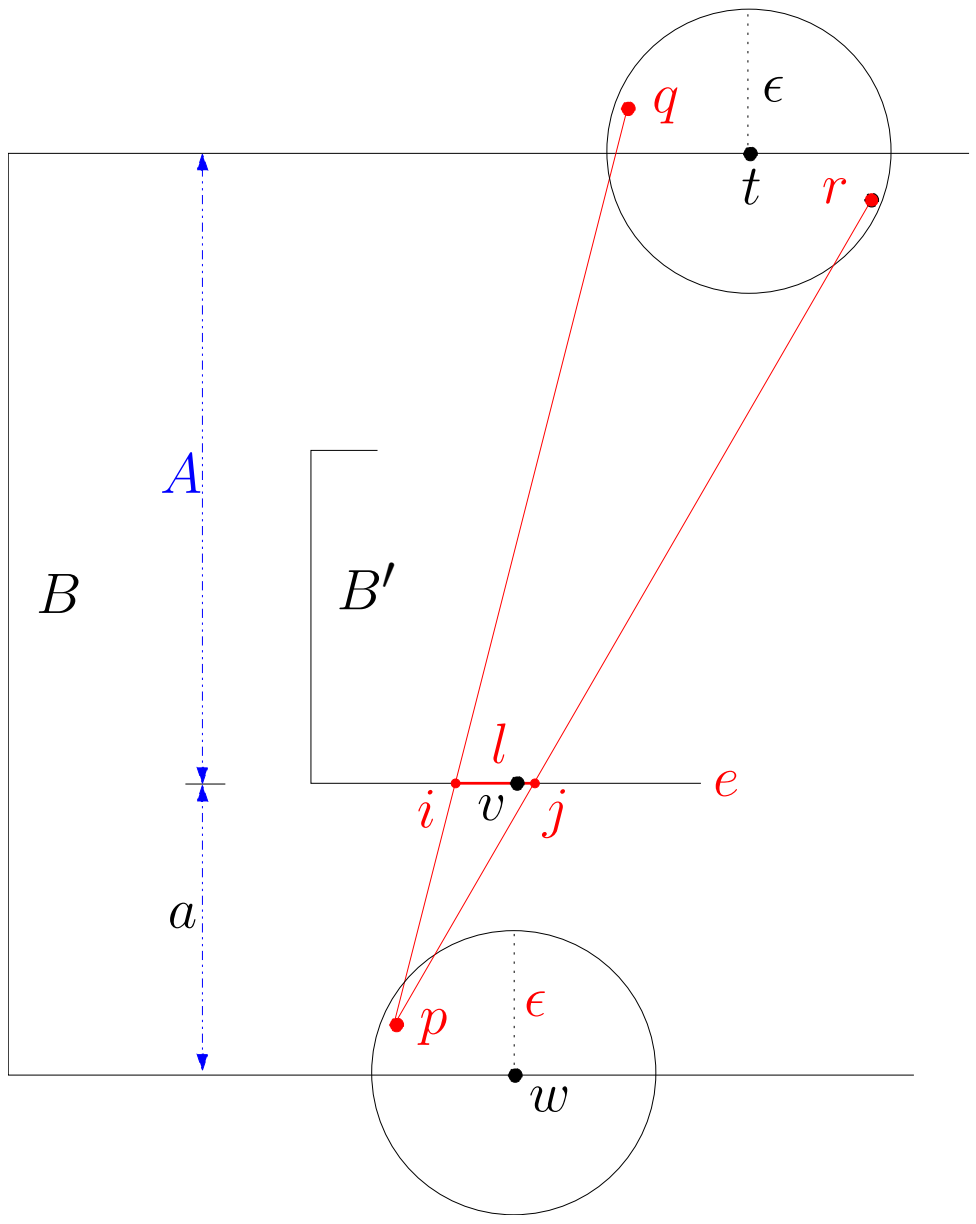

Figure 5: Including point $v$ in a wedge of segments that connect points of $V$.

Let $v$ be a point on the lower horizontal edge, $e$, of the smaller square $B'$, and let $w$ denote the point on $B$ below $v$; see Figure 5. Since $V$ is an $\epsilon$-cover for $B$, the $\epsilon$-neighborhood of $w$ contains some point $p$ of $V$. As we move a point $t$ from left to right along the upper edge of $B$, there is a position where the closure of $t$'s $\epsilon$-neighborhood contains two points, $q$ and $r$, of $V$ such that segment $pq$ crosses edge $e$ to the left of $v$, whereas $pr$ crosses to the right. By the law of rays, the distance $l$ between these crossing points, $i$ and $j$, roughly equals

$$l \approx \frac{a}{A+a}|qr| \quad \leq \quad \frac{a}{A+a}\, 2\epsilon. \tag{2}$$

The crucial fact is that, by choosing the parameters $a$ and $A$ appropriately, we can make sure that the factor $\frac{2a}{A+a}$ in (1) is smaller than the scaling factor $\frac{A-a}{A+a}$ of $B'$.

This observation is put to work in the following way. First, we include each of the four corners of square $B'$ in both a vertical and a horizontal wedge. The crossings of the wedge boundaries are points of $V^1$ and approximate $v$ to an error of—roughly—a constant times $l$, that is, more accurately than claimed in the lemma. This gives us some slack for approximating these crossing points by points of $V$, using assumption (2) of Lemma 3 and Lemma 2 with $k = 1$. If the value, $\delta$, of the dilation is chosen sufficiently small, we obtain four points $a, b, c, d$ of $V$ that approximate the corners of $B'$ to an error still smaller than the bound $\frac{A-a}{A+a} \cdot \epsilon$ we need. We draw the edges of the square $(a, b, c, d)$, and use them, instead of the ideal edges of $B'$, to form intersections with the wedge boundaries depicted in Figure 5. Again, the resulting crossing points are in $V^2$, and can thus be approximated by points of $V$ by the same argument as before. This shows that $V$ does indeed contain a set that is an $\frac{A-a}{A+a} \cdot \epsilon$-cover for $B'$, as desired. □

We observe that the values of $a, A, \epsilon, \delta$, for which Lemma 3 is valid, allow for simultaneous scaling of $a, A$, and $\epsilon$, while the dilation $\delta$ remains fixed.

**Theorem 5.** *Let $\epsilon$ and $\delta$ be values such that Lemma 3 is valid for $a, A, \epsilon, \delta$. If point set $Q$ is an $\epsilon$-cover for the boundary of a square of side length $A + a$ then $Q$ must have a dilation $\Delta(Q) > \delta$.*

*Proof.* If $\Delta(Q)$ were less than $\delta$ then there would be a triangulation $T$ with finite vertex set $V$ containing $Q$ such that $\delta(T) \leq \delta$. By Lemma 3, $V$ approximates, with greater precision $\frac{A-a}{A+a} \cdot \epsilon$, the boundary $B'$ of a square scaled down by the factor $\frac{A-a}{A+a}$, that is symmetrically included in $B$. The condition $\epsilon < a/2$ implies that the points of $V$ approximating $B'$ are different from those that approximate $B$. Since Lemma 3 can be applied ad infinitum, $V$ must be an infinite set—a contradiction. □

After working out the details of the proof of Lemma 3—which we do in the Appendix—one can easily verify that its claim is true for the values $a = 1, A = 15, \epsilon = 0.16$, and $\delta = 1.0000047$. Clearly, the condition on $Q$ in Theorem 5 can be easily met by the set of $\lceil \frac{4(A+a)}{2\epsilon} \rceil$ points evenly spaced along the boundary of a square of side length $A + a$. Thus, we obtain concrete examples like the following.

**Corollary 1.** *The set $Q$ of 200 points evenly spaced on the boundary of a square is of dilation $\Delta(Q) > 1.0000047$.*

## 5 Lower Dilation Bounds for General Point Sets

Now we are ready to prove our main result that was stated as Theorem 2 in the introduction.

**Theorem 2 (Gap Theorem).** *Every non-stabilizing finite point set in the plane has dilation strictly greater than 1.*

*Proof.* Let $P$ be a non-stabilizing point set. By Theorem 1, the limit $P^\infty$ of $P$ lies dense in some region $R$ of the plane. For some $L$, we can inscribe, in $R$, the boundary $B$ of a square of edge length $L$. Now let $a, A, \epsilon, \delta_1$, where $L = A + a$, be values for which Lemma 3 holds. There exists a finite order $k$ such that $P^k$ is an $\epsilon/2$-cover for $B$. By Lemma 2, there exists a $\delta_2$ such that for each triangulation $T$ of dilation $\delta(T) \leq \delta_2$, whose vertex set $V$ contains $P$, this set $V$ is actually an $\epsilon/2$-cover for $P^k$. Consequently, $V$ is also an $\epsilon$-cover for the boundary $B$, and Theorem 5 implies

$$\delta_1 < \Delta(V) \leq \delta(T) \leq \delta_2.$$

Consequently, there cannot be a triangulation $T$ of dilation $\leq \min(\delta_1, \delta_2)$ whose vertex set contains $P$, that is, $\Delta(P) > \min(\delta_1, \delta_2)$. □

## 6 Open Problems

We have shown that, with the exception of some special cases, finite point sets have a dilation strictly larger than 1, and we have provided an example set whose dilation is at least 1.0000047. How to compute the dilation $\Delta(P)$ of a given point set $P$ remains open. Even for the most simple case, the vertices of a regular 5-gon, the precise value is not known. The best embedding known is depicted in Figure 6.

Another interesting question is if the the value $\Delta(P)$ is always attained by a triangulation, or if there are cases where $\Delta(P)$ can only be obtained as a limit. If so, is there an upper bound to the weight of the approximating triangulations? Finally, it would be nice to have results on constructing good spanners, with or without Steiner points, if the number of crossings is bounded. Such results would not only be of theoretical interest; they would also be very useful for the design of real transportation networks.

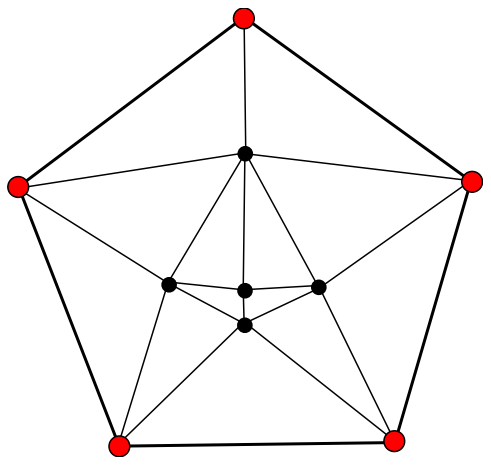

Figure 6: An embedding of five points in regular position into a triangulation of dilation 1.02046 due to Lorenz [15].

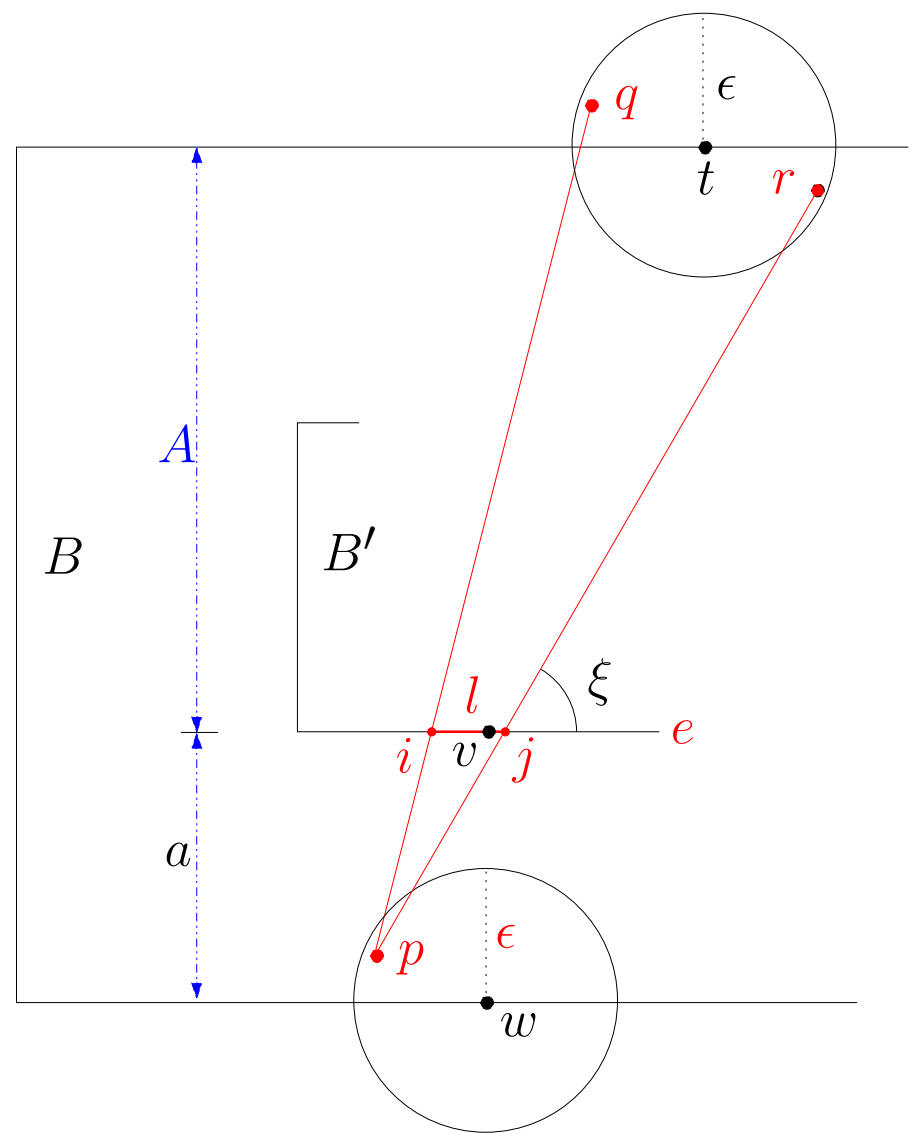

Figure 7: Including point $v$ between segments $pq$ and $pr$.

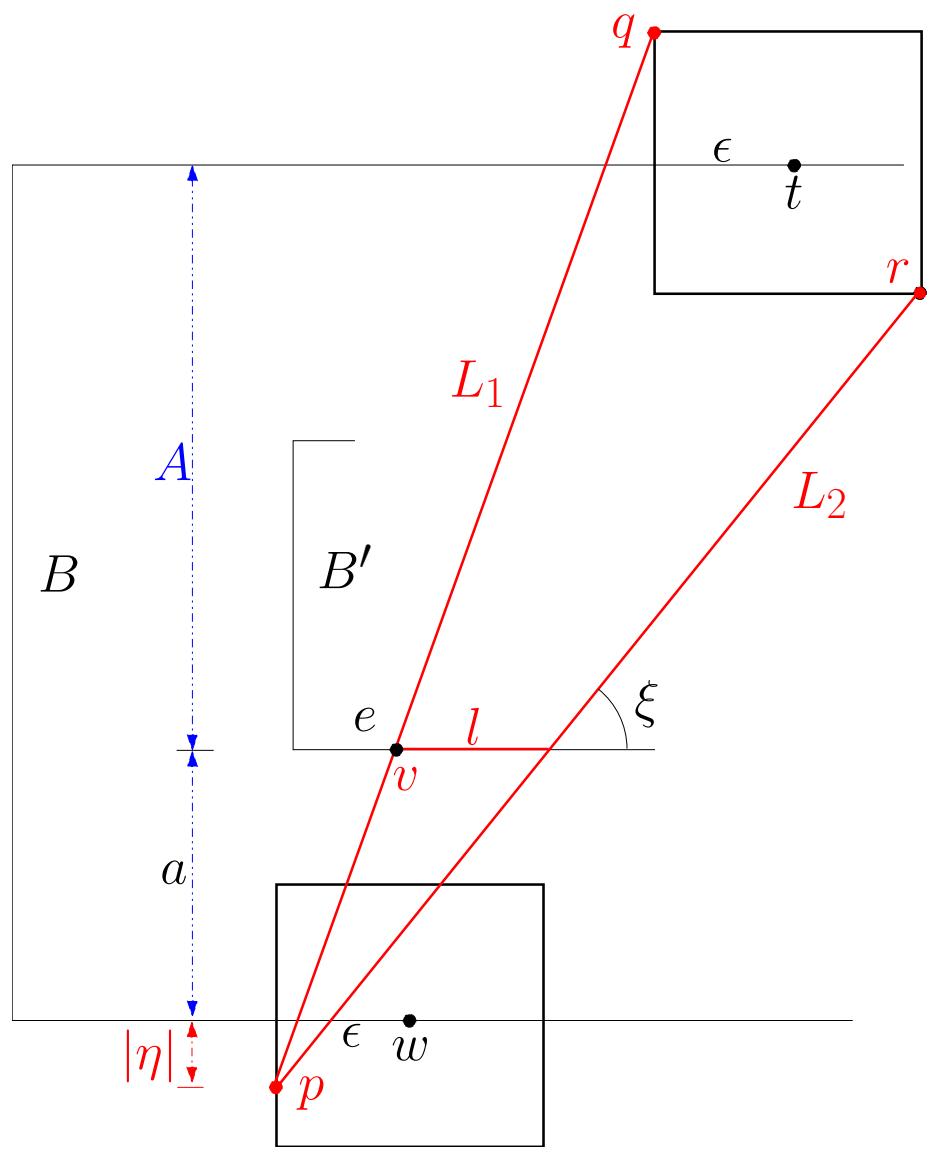

Figure 8: Considering bounding boxes.

# 7 Appendix

Here we provide the technical details of the proof of Lemma 3. They are necessary in computing concrete values of the parameters $a, A, \epsilon, \delta$ for which this lemma holds and, thus, for obtaining Corollary 1.

*Proof.* (**of Lemma 3**)

Let $v$ be a point on the lower horizontal edge, $e$, of $B'$, and let $w$ be the point on $B$ below $v$. Since $V$ is an $\epsilon$-cover for $B$, there is a point $p \in V$ in the $\epsilon$-neighborhood of $w$. Let $t \in V$ be a point on the upper horizontal edge of $B$ whose (closed) $\epsilon$-neighborhood contains two points $q, r$ of $V$ such that segment $pq$ passes to the left of $v$, and $pr$ passes to the right (as we move $t$ from left to right along the upper edge of $B'$ there must be a position where the switch occurs). We are interested in the maximum distance, $l$, between the intersection points $i, j$ of $pq$ and $pr$ with edge $e$, and in the smallest possible angle, $\xi$, between $pq, pr$ and the $X$-axis. To simplify computations, we include the $\epsilon$-neighborhoods in their bounding boxes. Also, we move $p$ to the left edge of the lower box, we place $q, r$ in extreme positions in the upper box, and we move the upper box to the right until $pq$ passes through $v$; see Figure 8. This can only increase $l$ and decrease $\xi$.

Let $v$ be the origin. Then the lines $L_1, L_2$ through $pq, pr$ are given by the equations

$$L_1(X) = \frac{a-\eta}{\epsilon}X \tag{3}$$

$$L_2(X) = \frac{(a-\eta)(A+a-\epsilon-\eta)}{\epsilon(A+3a+\epsilon-3\eta)}X - \frac{2(a-\eta)(a-\epsilon-\eta)}{A+3a+\epsilon-3\eta}, \tag{4}$$

where $\eta \in [-\epsilon, \epsilon]$ parametrizes the height of $p$ in its box. It turns out that the size of $l$ is maximized for $\eta = -\epsilon$, leading to

$$l = \frac{2\epsilon(a+2\epsilon)}{A+a} \tag{5}$$

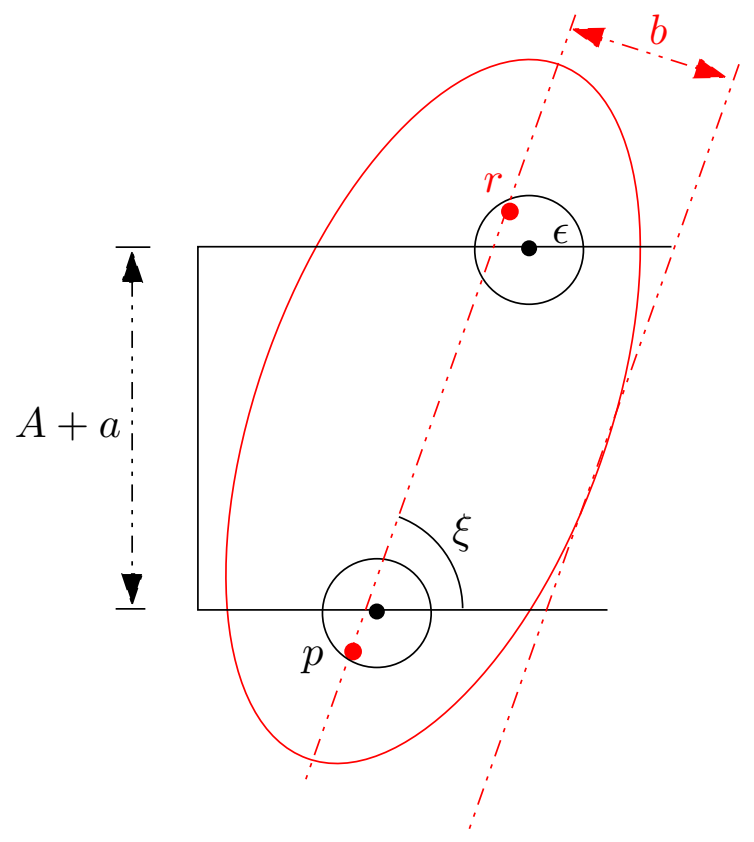

Figure 9: The ellipse with foci $p$ and $r$ and diameter $\delta|pr|$.

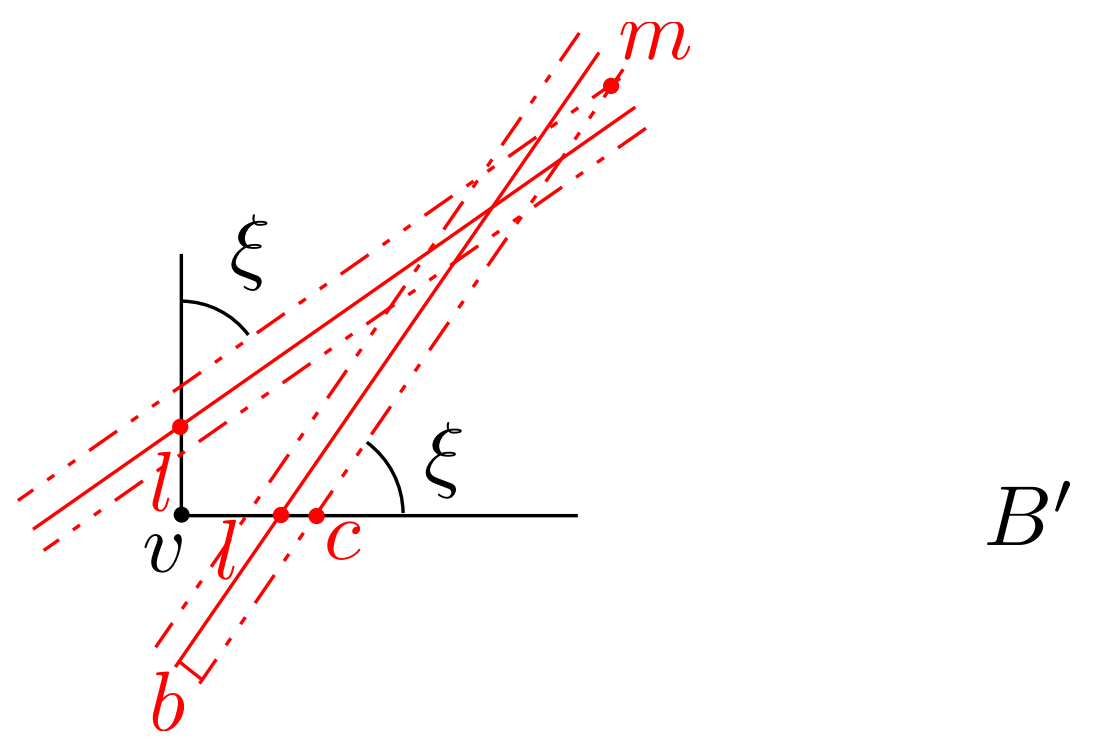

Figure 10: The closest point to $v$ in $V$ is at most distance $D = |vm|$ away from $v$.

whereas angle $\xi$ is minimized for $p$ in the upper left corner of its box, which results in

$$\tan\xi = \frac{(a-\epsilon)(A+a-2\epsilon)}{\epsilon(A+3a-2\epsilon)} \tag{6}$$

Next, we consider the pair of points $p, r$ of $V$. By assumption (2) of Lemma 3, there is a path of length $\leq \delta \cdot |pr|$ in triangulation $T$ that connects $p$ and $r$. This path must be completely contained in the ellipse of diameter $\delta|pr|$ with foci $p$ and $r$. Since $p, r$ are at distance at most $(A+a+2\epsilon)/\sin\xi$, the ellipse is of width at most

$$2 \cdot b \quad = \quad \frac{A+a+2\epsilon}{\sin\xi}\sqrt{\delta^2-1}; \tag{7}$$

see Figure 9. Now let $v$ denote the lower left corner of $B'$, and consider the vertical wedge including $v$. Arguing very generously, we know that within distance $l$ of $v$, an edge connecting two points of $V$ (namely $p, r$) of angle at least $\xi$ is passing through the lower edge of $B'$. Symmetrically, we can consider the horizontal wedge containing $v$; see Figure 10. If we surround each of these two segments by a strip of width $2b$ we can be sure that the closest point to $v$ in $V$ cannot be further away from $v$ than the furthest point, $m$, in the intersection of the strips. Applying the law of sines to the triangle with vertices $v, c, m$ depicted in Figure 10, we obtain for $D = |vm|$

$$\frac{D}{\sin\xi} \quad = \quad \frac{l + \frac{b}{\sin\xi}}{\sin(\xi - \frac{\pi}{4})}, \text{ hence} \tag{8}$$

$$D \quad = \quad (l + \frac{b}{\sin\xi})\frac{\sqrt{2}}{1-\cot\xi}. \tag{9}$$

Thus, we have obtained the following intermediate result.

**Lemma 4.** *$V$ is a $D$-cover for the four corners of the small square with boundary $B'$; here $D$ is given by formula (7).*

Finally, we approximate an arbitrary point $v$ on the lower edge of $B'$ in the following way. By Lemma 4 there exist points $d, d' \in V$ that approximate the lower corner of $B'$ to an error of $D$. Clearly, $|dd'| \leq A - a + 2D$, so that the ellipse with foci $d, d'$ of diameter $\delta|dd'|$ is of width at most

$$2b' \quad = \quad (A-a+2D)\sqrt{\delta^2-1}. \tag{10}$$

We consider the intersection of the segment $dd'$ with the least steepest edge, $pr$, of the vertical wedge depicted in Figure 8. Again, a point of $V$ must be contained in the intersection of the box of width $2b'$ around segment $dd'$ with the box of width $2b$ centered at $pr$; see Figure 11. Let $y$ denote the furthest point to $v$ in the strip intersection. Its vertical height over edge $e$ equals $D + b'$; hence, $|zy| = (D + b')/\sin\xi$, and from the law of cosine we obtain

$$F := |vy| \leq \sqrt{(l + \frac{b}{\sin\xi})^2 + (\frac{D + b'}{\sin\xi})^2 + 2(l + \frac{b}{\sin\xi})(\frac{D + b'}{\sin\xi})\cos\xi} \tag{11}$$

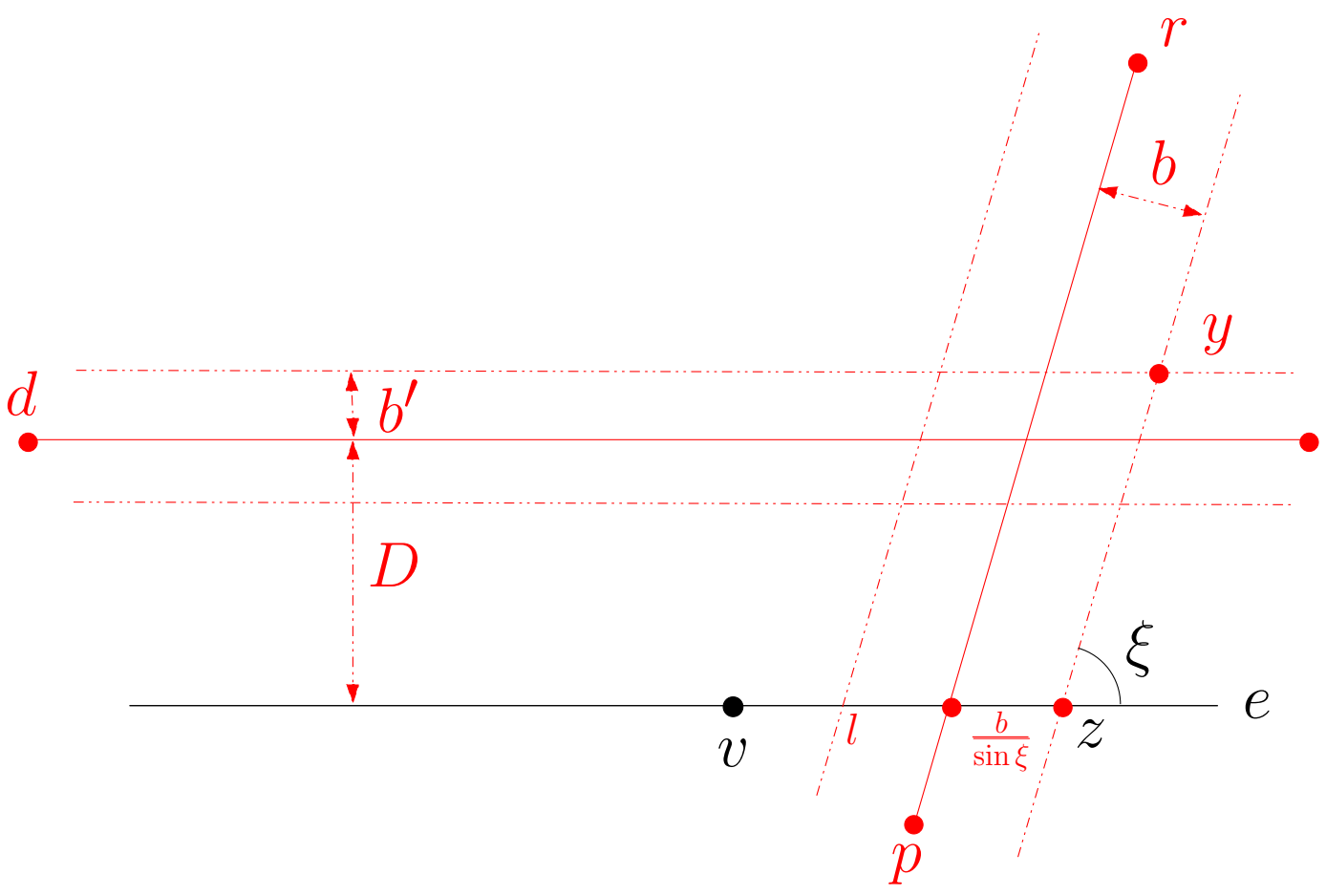

Figure 11: The distance from $v$ to the closest point in $V$ is at most $|vy|$.

Using the previous equations, the values of $F$ can now be numerically evaluated, and shown to be less than $\epsilon' := \frac{A-a}{A+a}\epsilon$ for suitably chosen parameters. For example, for $a = 1, A = 15, \epsilon = 0.16$ and $\delta = 1.0000047$ we obtain $F = 0.139904$, which is less than $\epsilon' = 0.14$. This completes the proof of Lemma 3. $\square$